\documentclass[twocolumn,twoside,slac_two]{revtex4-1}
\usepackage{graphicx}
\usepackage{amsmath}
\usepackage{fancyhdr}
\usepackage{xspace}
\usepackage{natbib}

\newcommand{\fu}{4U\,1909+07\xspace}

\newcommand{\inte}{\textsl{INTEGRAL}\xspace}

\newcommand{\xte}{\textsl{RXTE}\xspace}

\begin{document}

\title{\fu: a well-hidden pearl}

%

\author{F. F\"urst, I. Kreykenbohm, L. Barrag\'an, J. Wilms }
\affiliation{Remeis Observatory \& ECAP, Sternwartstr.~7, 96049~Bamberg, Germany }
\author{R. E. Rothschild, S. Suchy}
\affiliation{CASS, UCSD, 9500 Gilman Drive, La Jolla, CA 92093, USA}
\author{K. Pottschmidt}
\affiliation{CRESST and GSFC, Code 661, Greenbelt, MD 20771, USA and\\ 
UMBC, 1000 Hilltop Circle, Baltimore, MD 21250, USA}

\begin{abstract}
We present the first detailed spectral and timing analysis of the High Mass X-ray Binary (HMXB) \fu with \inte and \xte. \fu is detected with an average of 2.4 cps in ISGRI, but shows flares up to $\sim$50\,cps. The system shows a pulse period of 605\,s, but we found that the period changes erratically around this value. The pulse profile is extremely energy dependent: while it shows a double peaked structure at low energies, the secondary pulse decreases rapidly with increasing energy and above 20\,keV only the primary pulse is visible.
This evolution is consistent between PCA, HEXTE and ISGRI. 
We find that the phase averaged spectrum can be well fitted with a photoabsorbed power law with a cutoff at high energies and a blackbody component. To investigate the peculiar pulse profile, we performed phase resolved spectral analysis. We find that a change in the cutoff energy is required to fit the changing spectrum of the different pulse phases. 

\end{abstract}

\maketitle

\thispagestyle{fancy}


\section{Introduction}
The X-ray source \fu (or X1908+07) was detected with the \textsl{Uhuru} satellite \citep{giacconi74a} as 3U~1912+07 and filed in the fourth \textsl{Uhuru} catalog as \fu \citep{forman78a}. It has been seen with most X-ray instruments ever since. The exact nature of the system was, however, not clear for almost 30 years. In 2000, \citet{wen00a} analyzed \xte ASM data and found a stable 4.4\,days period, which was interpreted as the orbital period of a binary orbit. Due to the high absorption, however, no optical counterpart could be identified.  Using \xte-PCA data, \citet{levine04a} found a second, shorter period of $\sim$605\,s in the X-ray flux, explained as the pulse period of a slowly rotating neutron star. Using Doppler delays curve \citet{levine04a} could refine the binary orbit parameters. They estimated the mass of the companion star to be $M_\star = 9$--$31$\,M$_\odot$ using a canonical mass of 1.4\,M$_\odot$ for the neutron star and the radius of the companion  star to be $R_\star \leq 22\,R_\odot$. One year later \citet{morel05a} detected an OB star in the near infrared at the location of the X-ray source, thus confirming that the system is a High Mass X-ray Binary (HMXB). The distance of the system was estimated to be 7\,kpc \citep{morel05a}. Prior to this discovery, \citet{levine04a} argued that the companion star could be a Wolf-Rayet star, which would make this system a possible progenitor system to a neutron star-black hole binary. This possibility, although intriguing, can be ruled out after the discovery of \citet{morel05a}.
Although the system shows no eclipse the X-ray flux is still strongly orbital dependent, as seen in ASM data folded onto the orbital period \citep{levine04a}. \citet{levine04a} also analyzed orbital resolved PCA spectra and found that around orbital phase 1 the photoabsorption increases dramatically by a factor of at least 3, explaining the decreased ASM flux. This increase in absorption can be very well described by a spherical wind model and an inclination of $54^\circ\leq i \leq 70^\circ$, depending on the parameters of the wind model.
Like most wind-fed HMXB with a neutron star, \fu shows a constant, irregular flaring behavior, superimposed on the regular pulsations. These flares are most probably due to inhomogeneities in the wind of the donor star, which lead to different accretion rates.
In these proceedings we present the data and reduction methods in Sect.~\ref{sec:data}. In Sect.~\ref{sec:timing} we analyze the pulse period evolution and the pulse profiles and in Sect.~\ref{sec:spectra} we perform phase averaged and phase resolved spectroscopy. We summarize and discuss our results in Sect.~\ref{sec:disc}.

\section{Data and Observations}
\label{sec:data}
For our study of the \fu system we used all available public data of the X-ray satellites \inte \citep{winkler03a} and \xte. \inte features the detector ISGRI, which is a coded mask instrument sensitive in the 15\,keV -- 10\,MeV energy range and part of the  Imager on Board the Integral Satellite \citep[IBIS;][]{ubertini03a,lebrun03a}. 
Even though no pointed observations on \fu were performed with \inte, thanks to ISGRI's large field of view of almost $30^\circ\times30^\circ$ more than 6\,Msec of off-axis data exist. A huge part stems from the core program on the galactic center performed in the early AOs \citep{winkler01a}. Data from later AOs was taken from the extended monitoring campaign of the microquasar GRS~1915$+$105 \citep{rodriguez08a}.
Twelve sources with a detection significance $> 6\sigma $ were found in an ISGRI mosaic consisting of the images of all ScWs, with \fu being the fourth brightest source. The data were extracted using the standard pipeline of the Offline Science Analysis (OSA) 7.0 for spectra and images and using \texttt{ii\_light} to obtain lightcurves with a higher temporal resolution. We extracted the lightcurve with a resolution of 20\,sec as a trade-off between good signal-to-nose ratio and good time resolution to measure the pulse period.

For phase resolved spectroscopy we used \xte observations carried out in early 2003 which were part of the analysis done by \citet{levine04a}. These data had a total exposure of 196\,ksec. Both major instruments of \xte were fully operational at that time, the Proportional Counter Array \citep[PCA;][]{jahoda06a}, sensitive in the 2--60keV energy band and the High Energy X-Ray Timing Experiment \citep[HEXTE;][]{rothschild98a}, sensitive between 15--250\,keV. Both instruments have a large effective area and a high temporal resolution, but no imaging capabilities.
\xte data were reduced using the standard HEASARC software (v.~6.6), obtaining lightcurves of the PCA with 1\,sec resolution and pulse phase resolved spectra for PCA and HEXTE in the 4.5--150\,keV range.

\section{Timing Analysis}
\label{sec:timing}

\subsection{Pulse period evolution}
\begin{figure}
 \centering
 \includegraphics[width=0.45\textwidth]{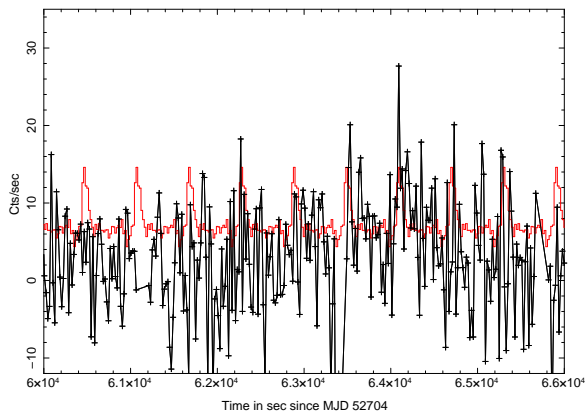}
 \caption{Small part of the lightcurve in the 20--40\,keV energy band as measured with ISGRI. The red line shows the folded pulse profile over the whole data set. }
 \label{fig:lc2pprof}
\end{figure}

As the pulse period of the system was only discovered in 2004 using \xte data from 2000 \citep{levine04a}, no historic data on the pulse period evolution exist. Between 2000 and 2004, the pulse stayed at a value of $P \approx 605.68$\,s \citep{levine04a}. The archival \inte data now provide an optimal basis to perform many more measurements of the pulse period between 2003 and 2008 and to follow its evolution. For this analysis we splitted the lightcurve into segments between 300\,ksec  and 800\,ksec length, providing a good balance between accurate pulse period determination and temporal resolution, and used epoch folding \citep{leahy87a} to determine the pulse period for each of these segments individually. \fu shows strong pulse-to-pulse variations, in fact, the pulse period is not visible in the ISGRI lightcurves to the naked eye, see, e.g., Fig. \ref{fig:lc2pprof}. Only after folding 500--1000 pulse periods, a stable pulse profile arises which allows for a reliable pulse period determination. The emerging profile is, however, remarkably stable over our data set. A similar effect is seen in Vela~X-1, although the pulsed fraction and the overall luminosity are distinctly higher in that source \citep{staubert80a}.

\begin{figure}
 \centering
 \includegraphics[width=0.45\textwidth]{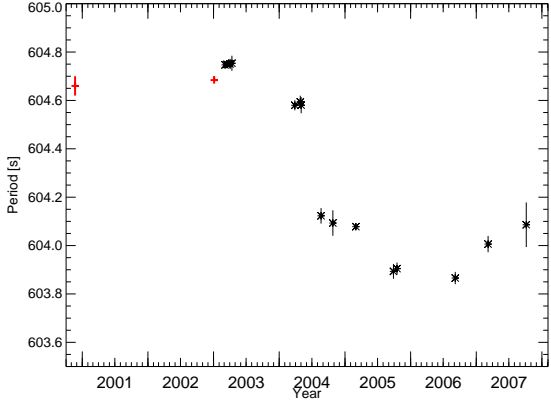}
 \caption{Evolution of the pulse period over the last years. The historic \xte data points of \citet{levine04a} are shown as red crosses, the new measurements obtained with \inte are shown as stars. }
 \label{fig:per_evolu}
\end{figure}

\begin{figure}
 \centering
 \includegraphics[width=0.45\textwidth]{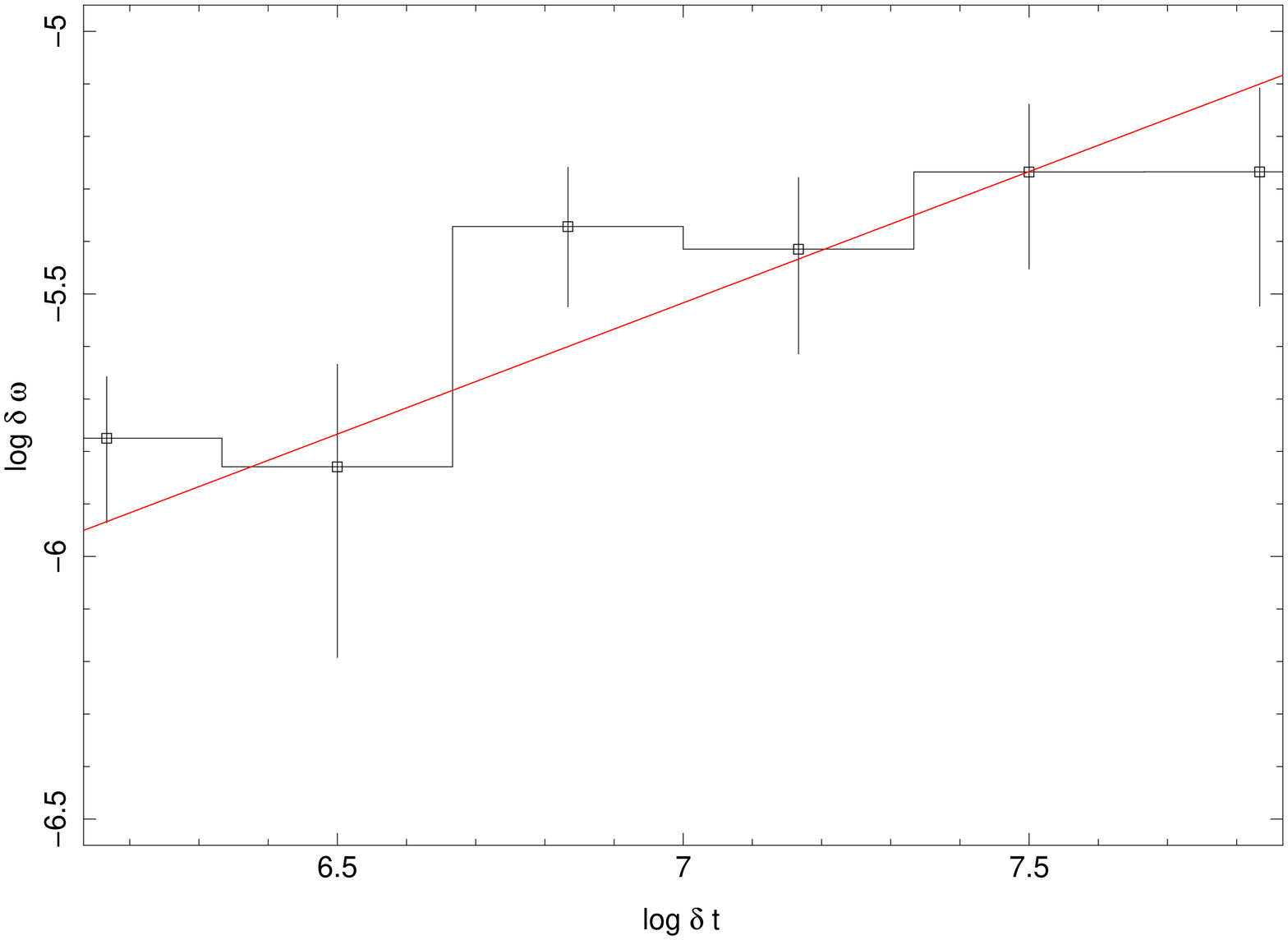}
 \caption{Plot of the pulse period evolution in $\log \delta \omega - \log \delta t$ space, as proposed by \citet{dekool93a}. A ideal random walk process would show a slope of 0.5 in this plot. The data follow this slope closely, as shown by the red straight line with a slope of 0.5.}
 \label{fig:logom_logt}
\end{figure}

The pulse period between 2001 and 2007 is shown in Fig.~\ref{fig:per_evolu}. Albeit two trends could be picked out, a spin-up trend between March 2003 and April 2006  and a spin-down trend from April 2006 to October 2007 the overall behavior is consistent with a random walk like behavior. To confirm this observation, we implemented the algorithm proposed by \citet{dekool93a}. This algorithm evaluates the relative change in pulse period over different time intervals $\delta t$ between single measurements. For a perfect random walk, the result will be a straight line with a slope of 0.5 in $\log \delta \omega - \log \delta t$ space, where $\omega$ is the angular velocity: $\omega = {2 \pi}/{P}$. As only very few measurements of the pulse period of \fu were available, the outcome of the calculation is relatively coarse. Nonetheless, it is clear from Fig. \ref{fig:logom_logt} that the data follow a straight line. Superimposed on the plot is a line with slope of 0.5, shifted in $y$-direction to fit the data. The uncertainties to the data are not only due to uncertainties in the estimate of the pulse period, but are also due to the uneven and coarse sampling. To estimate this effect, we choose a Monte Carlo approach and simulated many mock-up pulse period evolutions which followed a perfect random walk and sampled these with the same rate as the real data \citep[see][]{dekool93a}. The standard deviation in each bin from these simulations gives an estimate of the uncertainty in each bin and is roughly of the same order of magnitude as the uncertainties from the pulse period determination.

\subsection{Pulsprofiles}
\begin{figure}
 \centering
 \includegraphics[width=0.45\textwidth]{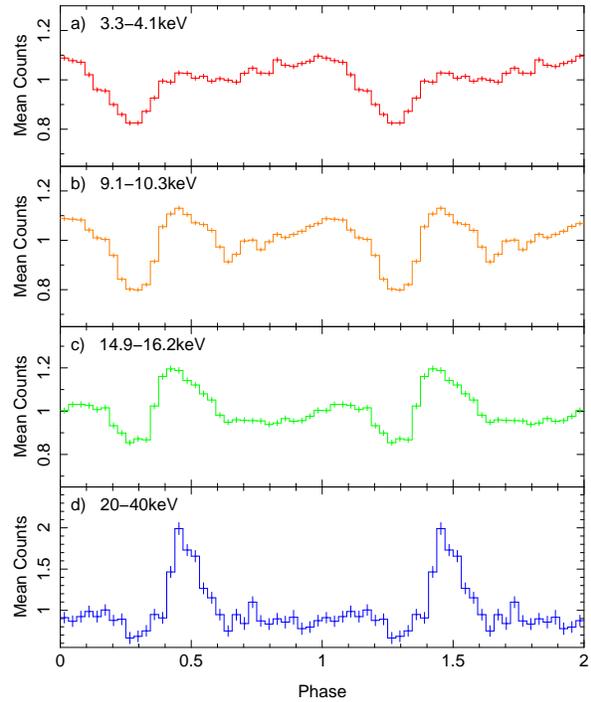}
 \caption{Energy resolved pulse profiles with \xte PCA (\textit{a)--c)}) and \inte ISGRI (\textit{d)}). The profiles are shown twice for clarity. Note that the \xte and \inte profiles are not phase aligned.}
 \label{fig:pp_energy}
\end{figure}

As shown by \citet{levine04a}, the  3.7 -- 17\,keV pulse profile shows two distinct peaks, with the second one being slightly broader than the first one and having a complex shape with two subpeaks at phase $\phi =0.85$ and $\phi=1.1$. 
But the pulse profile of \fu shows a remarkable and strong energy dependence. 
To follow this energy dependence of the pulse profile, we extracted pulse profiles from PCA data in different energy bands, using a period of $P = 604.685$\,s and 32 phase bins. Thanks to the large effective area of the instrument, very good pulse profiles could be extracted in 30 narrow energy bands. Three examples are shown in the upper panels of Fig.~\ref{fig:pp_energy}. The pulse profiles show a smooth transition from a two peaked profile at low energies to a single peak profile at high energies. At very low frequencies the secondary peak is broader and stronger than the first one (Fig.~\ref{fig:pp_energy}a). This behavior levels out with increasing energy and the two peaks become more separated (Fig.~\ref{fig:pp_energy}b). The relative power of the secondary peak keeps declining with increasing energy and is very small compared to the first one at energies around 20\,keV (Fig.~\ref{fig:pp_energy}c). The deep minimum around phase 0.3 is not energy dependent. For the very high energies we extracted the pulse profile from \inte data of the average data of the first 100 days of measurement in 2003. We used the same epoch as for the PCA analysis, but a period of $P = 604.747$\,s, as determined in our analysis of the data. The trend seen in \xte pulse profiles is continued, and the secondary peak has completely vanished, where as the primary is very clearly seen (Fig.~\ref{fig:pp_energy}d).

\section{Spectral analysis}
\label{sec:spectra}
\subsection{Phase averaged spectrum}
\begin{figure}
 \centering
 \includegraphics[width=0.4\textwidth]{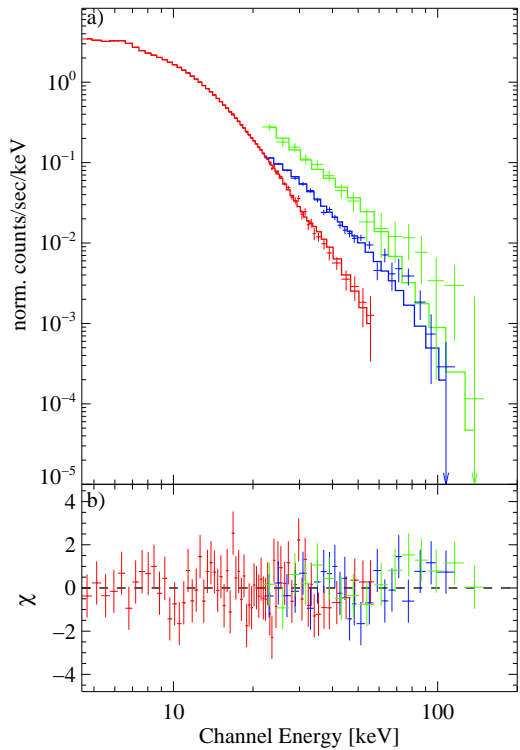}
 \caption{\textit{a)} Combined \xte PCA (red) and HEXTE (blue) and \inte ISGRI (green) spectra. The best fit \texttt{cutoffpl}-model is also shown. \textit{b)} Residuals of the best fit model.}
 \label{fig:spec_low}
\end{figure}

The pulse phased averaged spectrum of \fu shows a shape similar to many other neutron star sources, consisting of a powerlaw continuum attenuated by photoabsorption at low energies and by a turnover at high energies. \citet{levine04a} modelled the spectrum using a cutoff-powerlaw and a bremsstrahlung model and found that both fit the data equally well. The turnover at high energies is modeled phenomenologically with different models like \texttt{cutoffpl} or \texttt{highecut}, where in the latter the cutoff is only applied after a certain energy $E_\text{cut}$. More sophisticated ones include \texttt{fdcut} \citep{tanaka86a} and \texttt{npex} \citep{mihara95a}. 
For our spectral analysis we discarded all observations between orbital phase $0.88 < \phi_\text{orb} < 0.12$, where the $N_\text{H}$ is dramatically increased \citep{levine04a}.
To this dataset we fitted the above mentioned models to \xte and \inte data simultaneously using XSPEC 11.3. We found that all models give equally good fits regarding the reduced $\chi^2$ value. In the \texttt{fdcut} model, however, the cutoff energy is fitted to values $<$1\,keV, thus effectively reducing the model to a \texttt{cutoffpl}. Some of the most important spectral parameters of all other models are shown in Tab.~\ref{tab:avgpara}. Independent of the applied continuum models, a soft excess below 10\,keV is evident, which can be very well modelled using a blackbody with a temperature of $kT\approx$1.4\,keV. Additionally strong evidence for an iron line close to the energy of neutral iron of 6.4\,keV is found. The iron line was modelled with a Gaussian with an equivalent width of $\sim$250\,eV.  Fig. \ref{fig:spec_low} shows the spectrum and the best fit \texttt{cutoffpl} with its residuals. 

\begin{table*}
\caption{Fit parameters for the phase averaged spectra with different models}
\label{tab:avgpara}
\centering
\begin{tabular}{ccccccc}
\hline\hline
Model     & Cutoffpl & Cutoffpl & Highecut & Highecut & NPEX & NPEX \\
parameter &          & +bbody   &          &   +bbody       &      &  +bbody\\\hline
$\chi^2_\text{red}$ & 1.76 & 1.01 & 1.07 & 0.91 & 1.44 & 0.93 \\
$N_\text{H}$ [atoms\,cm$^{-2}$] & $15.3^{+0.6}_{-0.5}$ & $4.7^{+1.6}_{-1.9}$ & $4.8^{+0.9}_{-1.6}$ & $6.8^{+2.0}_{-0.8}$ & $1.7^{+0.7}_{-1.7}$ & $5^{+1}_{-3}$ \\
$\Gamma$\footnote{For the NPEX models $\alpha_1$ and $\alpha_2$ are given} & $1.63^{+0.13}_{-0.02}$ & $0.96^{+0.03}_{-0.06}$ & $1.37^{+0.03}_{-0.08}$ & $1.32\pm0.10$ & $0.36^{+0.05}_{-0.15}$ / $-3.08^{+0.13}_{-0.09}$ & $0.80^{+0.06}_{-0.15}$ /  -2.0\footnote{$\alpha_2$ was frozen during fitting}\\
Fe $\sigma$ & $1.7^{+0.1}_{-0.5}$ & $0.28^{+0.15}_{-0.11}$ & $0.41^{+0.07}_{-0.06}$ & $0.2\pm0.2$ & $0.41^{+0.07}_{-0.06}$ & $0.27^{+0.12}_{-0.18}$ \\
Fe Energy [keV] & $6.4^{+0.8}_{-0.2}$ & $6.40^{+0.04}_{-0.06}$ & $6.39^{+0.04}_{-0.03}$ & $6.39^{+0.04}_{-0.05}$ & $6.37^{+0.04}_{-0.03}$ & $6.40\pm0.05$ \\
Fe Norm & $(1.1^{+0.1}_{-0.6})\times10^{-3}$ & $(0.57^{+0.17}_{-0.11})\times10^{-3}$ & $(0.78^{+0.07}_{-0.04})\times10^{-3}$ & $(5^{+3}_{-1})\times10^{-4}$ & $(0.80^{+0.07}_{-0.06})\times10^{-3}$ & $(0.6^{+0.3}_{-0.1})\times10^{-3}$ \\
\end{tabular} 
\end{table*}

\subsection{Phase resolved spectra}
\begin{figure}
 \centering
 \includegraphics[width=0.42\textwidth]{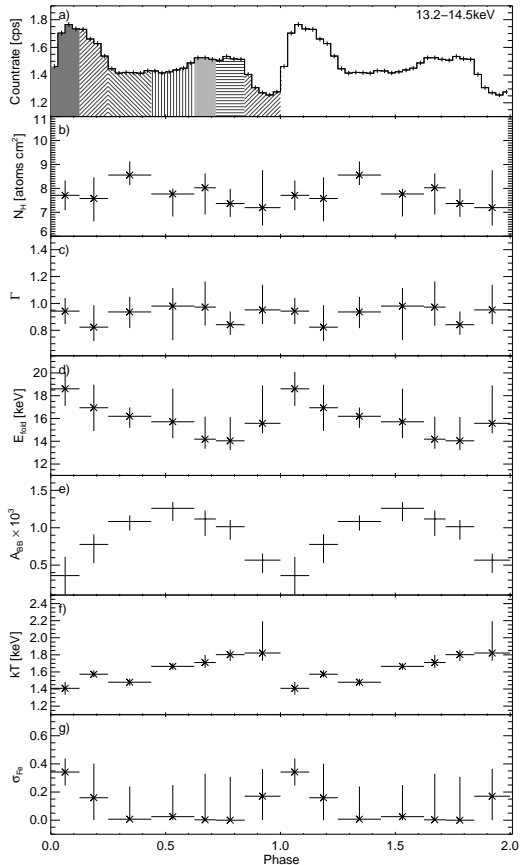}
 \caption{Parameters of the phase-resolved spectra fitted with the \texttt{cutoffpl} model. \textit{a)} Pulse profile in the 13.2-14.5\,keV energy range. The different shaded areas indicate the phasebins used to extract the spectra. \textit{b)} Photo electric absorption column $N_\text{H}$, \textit{c)} Power law index $\Gamma$, \textit{d)} Folding energy, \textit{e)} Black body norm, \textit{f)} Black body temperature, \textit{g)} Iron line width $\sigma$.}
 \label{fig:phasresparas}
\end{figure}

We divided the PCA and HEXTE data into 7 phasebins, as indicated in Fig.~\ref{fig:phasresparas}a). The \inte ISGRI data did not provide high enough statistics for high resolution phase resolved spectroscopy. To all phase bins we fitted a \texttt{cutoffpl} with black body and a \texttt{npex} model with black body. The \texttt{cutoffpl} and the \texttt{npex} models showed comparable fit quality and similar systematics. As the \texttt{cutoffpl} is the simpler model, we will only present these data here.
The evolution of the spectral parameters is shown in Fig. \ref{fig:phasresparas}. It is clearly seen that neither the photoabsorption $N_\text{H}$ nor the power law index $\Gamma$ seems to vary much over the pulse phase. The strong spectral changes indicated in the energy resolved pulse profiles are thus mainly due to a change in the cutoff energy of the \texttt{cutoffpl} model. The cutoff energy is highest in the primary peak, making its spectrum distinctly harder than the rest of the pulse phase. In the secondary pulse the cutoff energy moves to values as low as $\sim$14\,keV, forcing a strong attenuation of the spectrum at higher energies.
Very interesting is also the behavior of the black body component. Its temperature $kT$ is relatively stable over the pulse, but its normalization is changing by a factor of $\sim$3. The black body is strongest in the phase bin between the two peaks and lowest in the rise and maximum of the primary peak. This small contribution of the black body in the primary peak is somewhat compensated by a broadening of the iron line in that phasebin. With the energy resolution of the PCA it is impossible to tell  if the iron line really is stronger in that phasebin or if a mixture of blackbody and line is fitted.

\section{Discussion \& Conclusion}
\label{sec:disc} 
%
We have presented the first detailed study with \inte and \xte of \fu and have shown that the pulse period evolution shows strong indication for a random walk like behavior. Such a behavior has been seen in other HMXB like Vela~X-1 and is a strong indicator for a wind accreting source without a persistent accretion disk \citep{ghosh79a}. An accretion disk would provide a more constant transfer of angular momentum and thus a long-term spin-up or spin-down trend, as seen in other sources like 4U~1907+09 \citep{fritz06a}. Accretion disks form to fulfill the conservation of angular momentum of the accreted wind, but in a system with a strongly magnetized neutron star the ionized matter will couple to the magnetic field lines before a stable disk can form. This requires field strengths on the order of $10^{12}$\,G, not unusual for this kind of systems. In these strong magnetic fields, however, cyclotron resonant scattering features (CRSFs) should form, as seen in many other sources, e.g.,  Vela~X-1 \citep{kreykenbohm99a} or MXB~0656$-$072 \citep{mcbride06a}. We found no evidence for a CRSF in \fu. As shown by \citet{schoenherr07a} this does not rule out a strong magnetic field, as CRSFs can be filled up again by photon spawning and depend strongly on the geometry of the accretion column and on the viewing angle. Further investigations with high-resolution data will allow a more thorough search for a CRSF in \fu.
A strong magnetic field in \fu is also supported by the strongly pulsed flux, as the X-ray flux is believed to be produced in a very small region above the so-called hot-spots on the neutron star surface, where the magnetic field lines penetrate the neutron star surface \citep{lamb73a}. In the accretion column above the hot-spots the density and temperatures are highest and thus the most X-rays are produced through Comptonization of the thermal photons from the thermal mound on the neutron star surface \citep{becker05a, becker05b}. If the magnetic axis is not aligned with the rotational axis of the neutron stars, the hot spots and the accretion columns move in and out of sight of the observer, resulting in a pulsed X-ray flux.

The two distinct peaks in the pulse profile at low energies hint at the fact that accretion happens onto both magnetic poles of the neutron star, but under different physical conditions. The different conditions can explain the different spectra in the two peaks. The secondary peak shows a distinctly lower cutoff energy which corresponds roughly to the temperature of the electron gas in the accretion column. A possible explanation could be, e.g., different sizes of the hot-spots due to misalignment of the magnetic center and the gravitational center of the neutron star. This would lead to different sizes in the accretion column and thus to different densities and temperatures. 

The exact geometry of the accretion column depends strongly on the way the matter couples to the magnetic field and is highly uncertain \citep{basko76a, meszaros84c}. Besides the simple filled funnel other possible configurations include a hollow or a partly hollow funnel. With an average luminosity of $\sim 2.8\times 10^{36}$\,erg\,s$^{-1}$ the system can host accretion rates that are large enough for a shock to form in the accretion column, due to the radiation pressure from the material close to the neutron star surface. This shock prevents escape of the Comptonized photons parallel to the magnetic field lines, so that they can only escape on the sides of the accretion column in a so-called ``fan beam'' \citep{meszaros84c}. If the accretion rate is lower, no shock will form and the radiation can escape along the accretion column in the ``pencil beam''. Which of these two geometries exists in a given object is hard to tell from timing analysis alone, as no independent measurements of the alignment of the magnetic axis is possible. But phase-resolved spectroscopy can provide some insight into the configuration. For \fu we have seen that the blackbody component of the spectrum is phase shifted to the two peaks in the pulse profiles.  It is most likely that the blackbody originates from the thermal photons of the thermal mound at the hot spot. Through the rotation of the neutron star, the hot spot is only visible at certain pulse phases. Assuming a hollow funnel accretion geometry, with a large enough accretion rate to form a shock, we can argue that we view the thermal mound only at the phases were we look along the accretion column, while the most hard X-ray flux is only visible when we see the sides of the accretion column, due to the ``fan beam'' geometry. 

Caution has to be taken, however, with these models as the behavior of plasma in magnetic fields on the order of $B\approx 10^{12}$\,G is still only little understood and all models are based on gross simplifications. Additionally gravitational light bending must be taken into account when analyzing pulse profiles and their origin \citep{kraus03a}. Detailed analysis of this kind is ongoing and will be presented in a separate publication. 

\bigskip 
\begin{acknowledgments}
This work was supported by the Bundeministerium f\"ur Wirtschaft und Technologie through DLR grant 50\,OR\,0808 and via a DAAD fellowship. This work has been partially funded by the European Commission under the 7th Framework Program under contract ITN\,215212. FF thanks the colleagues at UCSD and GSFC for their hospitality. This research has made use of NASA's Astrophysics Data System. This work is based on observations with \inte, an ESA project with instruments and science data centre funded by ESA member states (especially the PI countries: Denmark, France, Germany, Italy, Switzerland, Spain), Czech Republic and Poland, and with the participation of Russia and the USA.
\end{acknowledgments}

\bigskip 


\end{document}